# Collective coupling of a macroscopic number of single-molecule magnets with a microwave cavity mode

A. W. Eddins[1♣], C. C. Beedle[2‡], D. N. Hendrickson[2], and Jonathan R. Friedman[1]*

[1]Department of Physics, Amherst College, Amherst, MA  01002

[2]Department of Chemistry & Biochemistry, University of California at San Diego, La Jolla, CA 92093

We report the observation of strong coupling of a macroscopic ensemble of ~$10^{16}$ $Fe_8$ molecular nanomagnets to the resonant mode of a microwave cavity. We use millimeter-wave spectroscopy to measure the splitting of the system's resonant frequency induced by the coupling between the spins and the cavity mode. The magnitude of this splitting is found to scale with $\sqrt{N}$, where $N$ is the number of collectively coupled spins. We control $N$ by changing the system's temperature and, thereby, the populations of the relevant spin energy levels. Strong coupling is observed for two distinct transitions between spin energy states.  Our results indicate that at low temperatures nearly all of the spins in the sample couple with the cavity's resonant mode even though there is substantial inhomogeneous broadening of the $Fe_8$ spin resonances.

Single-molecule magnets (SMMs) are chemically synthesized materials in which each molecule behaves as an isolated nanomagnet. They have long been touted for their potential to become the highest density magnetic storage material, with one bit of information stored in each molecule[1], and there has been significant recent progress towards realizing this goal.[2] In tandem, because SMMs are quantum systems,[3] they have been suggested as possible qubits, the processing elements in quantum computers[4]. Quantum coherent phenomena have been observed in several SMMs.[5-8] Here we present evidence for a form of collective coherence in an SMM system in which the spins couple to the resonant mode of a microwave cavity. We find that nearly all of the ~ $10^{16}$ molecules in a crystal of the $Fe_8$ SMM collectively exchange photons with the cavity mode. The results suggest that SMMs may be used in a form of quantum magnetic storage in which information is stored holographically[9-14] in the entire crystal rather than bitwise in individual molecules.

Coherent coupling between two-level systems (e.g. spins) and cavity photons lies at the heart of cavity quantum electrodynamics. Such interactions have been seen in many systems, including individual atoms,[15] Bose-Einstein condensates,[16] semiconductor quantum dots,[17,18] and superconducting qubits.[19] Each of these systems couples to photons via electric-dipole transitions. Very recently, coupling cavity photons to spins via much weaker magnetic dipole transitions has been investigated.[20] This weaker coupling, while more challenging to observe, can lead to longer coherence times. Coupling of spins and cavity photons has now been observed in several low-spin systems, including standard electron-spin resonance materials,[21,22] nitrogen-vacancy centers in diamond,[13,23] $Cr^{3+}$ impurities in Ruby,[24] and N-doped buckyballs as well as a doped semiconductor.[12]

In all of these systems the spin belongs to a single atom, ion or nucleus. In contrast, SMMs are more "macroscopic" artificial magnets where the spin degree of freedom is a joint property of an entire metal-oxide molecular cluster. The macroscopic nature of these magnets also presents a complication: For many SMMs, variations in the local environment of the molecules within a crystal lead to slightly different properties for each molecule[25] and inhomogeneous broadening of spectral resonance lines[26]. At the same time, with ~ 1 molecule per unit cell and a large ($s \sim 10$) magnetic moment, SMMs also have an extremely high spin density, leading to a much stronger spin-photon interaction than what is seen in many other spin systems. Moreover, SMMs are in a regular crystalline array, which may increase the fidelity for the storage of quantum data. Our results show that the high spin density in SMMs can be harnessed to create a coupling strong enough to overcome the intrinsic inhomogeneity of the system.

The $Fe_8$ molecule (Fig. 1c) is a spin-10 object whose behavior can be well described by the spin Hamiltonian,

$$H = DS_z^2 + E(S_x^2 - S_y^2) + C(S_+^4 + S_-^4) - g\mu_B \vec{B}_{ext} \cdot \vec{S}, \qquad (1)$$

where $\vec{B}_{ext}$ is the externally applied magnetic field, $g \approx 2$, $D = -25.2$ μeV, $E = -4.02$ μeV, and $C = 7.4 \times 10^{-4}$ μeV[27]. The first term in Eq. 1 impels the spin to point parallel or antiparallel to the "easy" z-axis. This gives rise to a double-well potential, as shown in Fig. 1a. The spin has $2s + 1 = 21$ possible orientation states, $m = -10, -9, \ldots 10$. The zero-field energy difference between the $m = \pm 10$ and $\pm 9$ states corresponds to a frequency of ~ 114 GHz, while for the $m = \pm 9$ and $\pm 8$ states the energy difference corresponds to ~102 GHz. The component of a magnetic field parallel to the easy axis, $B_z = B_{ext}\cos\theta$, tilts the potential, as shown, and increases the energy differences between the lowest states.

Our electromagnetic cavity has distinct resonant modes; a specific cavity mode with $n$ photons is designated by $|n\rangle$. The coupling of a cavity mode to the transition between SMM energy levels will be appreciable only when the cavity frequency is near the frequency of the transition. An SMM's energy levels are typically anharmonically dispersed, as shown in Fig. 1a, and so only one pair of spin levels will couple effectively to the cavity at a time. Thus, the spin's energy-level spectrum can be truncated to these two levels, which behave as an effective spin-1/2 system. We relabel the lower of the relevant states as $|\uparrow\rangle$ and the higher one as $|\downarrow\rangle$, and define $\hbar\omega_S$ to be the energy difference between these two states. For example, when we truncate the states in Fig. 1a to the two lowest levels, we set $|\uparrow\rangle = |m = 10\rangle$ and $|\downarrow\rangle = |m = 9\rangle$.

When the resonant frequency of the cavity mode, $\omega_C$, differs significantly from $\omega_S$, the spin and the cavity are not appreciably coupled. This situation is represented in Fig. 1b by the dashed gray level labeled $|\downarrow\rangle$. In this limit, the energy states of the total system (spin and cavity) are well described by product states: $|\uparrow\rangle|n\rangle$ and $|\downarrow\rangle|n\rangle$. When the system has at most one excitation, the relevant basis states are $|\uparrow\rangle|0\rangle$, $|\uparrow\rangle|1\rangle$, and $|\downarrow\rangle|0\rangle$, which correspond to, respectively, the ground state of both systems, a photon in the cavity mode, and the excitation of the spin. The two systems can be coupled by applying an external magnetic field, which increases $\omega_S$, raising the energy of the $|\downarrow\rangle$ state (from dashed to solid level in Fig. 1b). The lower dashed red line in Figs. 2b,c shows the dependence of $\omega_S$ on field for the $m = 10$-to-9 transition. When $\omega_S$ becomes close to $\omega_C$ (vertical dashed black line in Figs. 2b,c), the spin will interact with the cavity mode by absorbing and emitting a photon. In this regime, the system's excited states hybridize, resulting in two split energy states, as illustrated in Fig. 1b.

Such a coupled system can be modeled by the Jaynes-Cummings Hamiltonian[28]:

$$H = H_{spin} + H_{rad} + H_{int}, \qquad (2)$$

where $H_{spin} = \frac{\hbar\omega_S}{2}(1-\sigma_z)$ is the spin Hamiltonian (i.e. Eq. 1 truncated to the two relevant levels), $H_{rad} = \hbar\omega_C n = \hbar\omega_C a^\dagger a$ is the Hamiltonian for the cavity mode, and $H_{int} = \frac{\hbar g_1}{2}(a\sigma_- + a^\dagger \sigma_+)$ is the Hamiltonian for the spin-photon interaction in the rotating-wave approximation. The $\sigma$'s are the standard Pauli spin matrices applied to the {|↑>, |↓>} basis and $a^\dagger$ ($a$) is the photon creation (annihilation) operator for the cavity mode. Offsets have been chosen to make the energy of the |↑>|0> ground state zero. The spin-radiation interaction strength, $g_1$, is given by

$$g_1 = |\langle\uparrow|S_T|\downarrow\rangle| g\mu_B B_{rf}/\hbar, \quad (3)$$

where $B_{rf}$ is the radiative magnetic field of a single cavity photon and $S_T$ is the projection of the spin operator in the direction of $B_{rf}$. The subscript "1" in $g_1$ refers to the fact that a single spin is coupled to the cavity. With $n \leq 1$, the excited eigenstates of Eq. 2 are:

$$|+\rangle = \sin(\varphi/2)|\downarrow\rangle|0\rangle + \cos(\varphi/2)|\uparrow\rangle|1\rangle$$
$$|-\rangle = \cos(\varphi/2)|\downarrow\rangle|0\rangle - \sin(\varphi/2)|\uparrow\rangle|1\rangle$$

with energies

$$E_\pm = \frac{\hbar}{2}\left((\omega_C + \omega_S) \pm \sqrt{\Delta^2 + 4g_1^2}\right), \quad (4)$$

where $\Delta = \omega_C - \omega_S$ is the cavity-spin detuning and $\tan\varphi = \frac{2g_1}{\Delta}$. Eq. 4 describes two branches of a hyperbola with asymptotes $E = \hbar\omega_C$ and $E = \hbar\omega_S$. For large $\Delta$, the excited eigenstates approach the independent excitation of the cavity or the spin, respectively. When $\Delta = 0$ the splitting between the two branches, $E_+ - E_-$, is $2\hbar g_1$, a quantity sometimes referred to as the vacuum-Rabi splitting, and the excited states of the system become simply $|\pm\rangle = \frac{1}{\sqrt{2}}(|\downarrow\rangle|0\rangle \pm |\uparrow\rangle|1\rangle)$, the two split states in Fig 1b.

Using the structure of our cavity mode, it is straightforward to calculate the single-photon field at the position of the sample to be $B_{rf} = 3.7(6) \times 10^{-7}$ G. Eq. 3 then yields $g_1/2\pi = 2.4$ Hz, much too small to be detected in a realistic experiment.

The situation changes dramatically when a large number of spins collectively couple to the cavity. When $N$ spins are contained within a volume much smaller than the photon wavelength, it is impossible to determine which spin absorbs or emits a photon. The two coupled spin-photon states then become

$$|+\rangle_N = \sin(\varphi/2)|\Downarrow\rangle|0\rangle + \cos(\varphi/2)|\Uparrow\rangle|1\rangle$$
$$|-\rangle_N = \cos(\varphi/2)|\Downarrow\rangle|0\rangle - \sin(\varphi/2)|\Uparrow\rangle|1\rangle \quad , \quad (5)$$

where

$$|\Uparrow\rangle = |\uparrow\uparrow\uparrow\ldots\uparrow\rangle$$

describes $N$ spins in the ground state and

$$|\Downarrow\rangle = \frac{1}{\sqrt{N}}\left(|\downarrow\uparrow\uparrow\ldots\uparrow\rangle + |\uparrow\downarrow\uparrow\ldots\uparrow\rangle + |\uparrow\uparrow\downarrow\ldots\uparrow\rangle + \ldots + |\uparrow\uparrow\uparrow\ldots\downarrow\rangle\right)$$

describes an equal superposition of each spin being flipped into the excited state (while the remainder stay in the ground state).

As first shown by Tavis and Cummings[29], the interaction strength of $N$ identical spins to the cavity mode is

$$g_N = \sqrt{N}\, g_1 \, . \quad (6)$$

For the case relevant to our experiments, where $N \sim 10^{16}$ spins couple to a cavity mode, $g_N/2\pi$ is $\sim 200$ MHz, an easily detectable frequency splitting.

In coupling to the cavity, the collection of $N$ spins behaves as one "superspin" with $s = N/2$[30]. The spin state $|\Downarrow\rangle$ corresponds to a rotation of the superspin vector by a small angle from the z axis (such that the z component of spin is reduced by 1). In our experiment the number of photons in the cavity $n$ is on the order of $10^{10}$. Nevertheless, Eq. 6 remains valid when the assumption $n \leq 1$ is replaced by the less stringent condition $n \ll N$. The latter corresponds to the limit in which the superspin's angle relative to the z axis remains small. The anharmonic limit, in which this angle is large, gives rise to superradiant states, as first noted by Dicke[30]. In practice, $N$ corresponds to the number of spins in the lower-energy state $|\uparrow\rangle$; $N$ depends on temperature and thereby permits *in situ* control of the coupling strength $g_N$.

Crystals of $Fe_8$ were synthesized using standard techniques. The crystal used for measurements was photographed under a microscope to determine its dimensions. Using those and the known unit cell for $Fe_8$,[34] we determined that the sample consists of $N_0 = 2.3(4) \times 10^{16}$ SMMs.

Fig. 1d shows a photograph of a single crystal of the $Fe_8$ SMM mounted in our cylindrical copper cavity. The $TE_{011}$ mode of our cavity has a resonance frequency of 147.677(2) GHz and Q ~ 4000. For this mode, the oscillating magnetic field, shown in Fig. 1e, is nearly perpendicular to the easy-axis. The sample is mounted such that its easy axis is $\theta$ ~ 35° from the external dc magnetic field, which is parallel to the cavity's symmetry axis. Our experimental set up is shown schematically in Fig. 2a. We performed measurements of the radiation power reflected from the cavity-sample system as a function of frequency and dc magnetic field at several temperatures between ~1.8 K and 20 K.

Figures 2b and 2c show absorbed power at 1.8 K and 7.0 K, respectively, for a range of frequencies and magnetic fields. Resonances of the system appear as yellow or red regions. The data exhibit two distinct resonant branches, each of which corresponds to one of the coupled spin-photon states in Eq. 5. At low magnetic fields, the resonances appear near the bare cavity resonance frequency (vertical dashed line) and the excitation frequency for the dipole-allowed $m$ = 10-to-9 spin transition (lower red dashed line). When the field approaches the value at which these resonances would cross in the absence of interaction, a clear avoided crossing opens up with the upper-left branch curving and eventually approaching the cavity resonance frequency. The lower right branch tends towards the spin transition frequency but signal strength is lost as the frequency increases. Irrespective of this loss of signal, we clearly see that there is a range of fields at which two resonance peaks are observed (see Supplementary Information, Fig. 2), a telltale sign the system is in the so-called strong coupling regime with states like those described by Eq. 5. Both branches can be fit very by Eq. 4 (with $g_1$ replaced by $g_N$) as shown by the black dashed curves in Fig. 2b. Only two parameters in the fit are unconstrained by the spin Hamiltonian or the cavity's resonant frequency: $\theta$, the angle between the easy axis and the magnetic field, a parameter that was restricted to be the same at all temperatures, and $g_N$, which was allowed to vary with temperature. The former determines the slope of spin transition frequency's field dependence (lower red dashed line) while the latter determines the gap between the two branches of the hyperbola. Our fits provided a best value of $\theta$ = 37.7°, close to the expected value of 35° based on the sample's orientation. For the data shown in Fig. 2b, we obtain $g_N/2\pi$ = 0.519(4) GHz.

Fig. 2b shows another, much smaller feature where the upper red dashed line, corresponding to the $m$ = 9-to-8 transition, intersects the cavity frequency; the feature is highlighted in the outset. Since at 1.8 K there are far fewer molecules in the excited $m$ = 9 state than in the $m$ = 10 ground state, the value of $N$ for the 9-to-8 transition is very small, resulting in a smaller coupling to the cavity mode. The splitting $g_N$ for this transition can be increased by raising the temperature, $T$, and thereby $N$ for the $m = 9$ state. Indeed, as shown in Fig. 2c, increasing $T$ to 7.0 K decreases the magnitude of the splitting associated with the (lower field) 10-to-9 transition and dramatically increases the coupling associated with the (higher field) 9-to-8 transition. This observation reflects the fact that raising the temperature monotonically reduces the population in the ground ($m = 10$) state while initially increasing the population of the excited ($m = 9$) state.

We fit the data in Fig. 2 (and similar data at other temperatures, not shown) to obtain values of $g_N$ for each spin-cavity resonance at all temperatures for which there was sufficient data to obtain a good fit to Eq. 4. Because $g_N = \sqrt{N}\left|\langle\uparrow|\vec{S}_T|\downarrow\rangle\right|\mu_B g B_{rf}/\hbar$, $g_N^2$ should be proportional to the relative population $p = N/N_0$ in the lower energy state of the relevant transition, where $N_0$ is the total number of spins in the sample. In Fig. 3, we plot

$$\left(\frac{\hbar g_N}{\left|\langle\uparrow|\vec{S}_T|\downarrow\rangle\right|\mu_B g}\right)^2 = pN_0 B_{rf}^2$$

as a function of temperature for the two transitions measured. It is straightforward to calculate the populations $p$ of the $m = 10$ and 9 states as a function of temperature with no adjustable parameters. The solid curves in Fig. 3 show this temperature dependence for the relevant levels, $m = 10$ and $m = 9$. The agreement between the data and the corresponding populations is striking. The only adjustable parameter for these curves is the product of $N_0$ and $B_{rf}^2$, which determines the vertical scale of the curves. Taking $N_0 = 2.3 \times 10^{16}$, we determine $B_{rf} = 5.30(1) \times 10^{-7}$ G for the 10-to-9 transition (Fig. 3a) and $5.03(3) \times 10^{-7}$ G for the 9-to-8 transition (Fig. 3b). These values agree well with each other and are on the same order as our calculated value of $3.7(6) \times 10^{-7}$ G using the structure of the $TE_{011}$ mode. The discrepancy may arise from modal mixing with the nearly degenerate $TM_{111}$ mode.

Inhomogeneous broadening in $Fe_8$, as in many SMMs, arises from variations within a sample of the anisotropy parameter $D$, as well as other Hamiltonian parameters.[26] The broadening can be seen in the rather wide spin resonances in the data in Fig. 2, which have a Gaussian width of ~ 760 Oe, corresponding to a frequency width of $\sigma_\omega/2\pi \sim 1.7$ GHz. A Gaussian distribution of $N$ spin resonant frequencies $\omega_s$ will still couple collectively if $\sigma_\omega \lesssim g_N$.[31] Our results do not quite meet this condition with $\sigma_\omega$ somewhat larger than $g_N$. The fact that we nevertheless observe collective coupling may indicate the existence of a small nonlinear coupling term in the spin-cavity interaction that induces the spins to synchronize[32], or the presence of some weak additional coupling mechanism, perhaps mediated by the crystal lattice[33]. These and other possible mechanisms are the subject of ongoing experimental and theoretical investigations. Regardless of the specific mechanism, our findings indicate that the spins need not have identical resonant frequencies in order to couple collectively to a cavity mode but can do so even with substantial inhomogeneous broadening.

We thank M. Bal for much useful advice and for his work on an earlier version of this experiment. We also thank E. M. Chudnovsky, J. A. Grover, D. S. Hall, D. B. Haviland, S. Hill, A. J. Millis, L. A. Orozco, M. P. Sarachik, D. I. Schuster, S. H. Strogatz, F. K. Wilhelm and D. J. Wineland for useful discussions and R. Cann for technical assistance and advice. Support for this work was provided by the National Science Foundation under grant nos. DMR-0449516 and DMR-1006519 and by the Amherst College Dean of Faculty.

| | |
|---|---|
| ♠ | Current address: Department of Physics, 366 LeConte Hall #7300, Berkeley, CA 94720 |
| ‡ | Current address: National High Magnetic Field Laboratory, 1800 E. Paul Dirac Drive, Tallahassee, FL 32310 |
| * | Corresponding author: jrfriedman@amherst.edu |

Figures

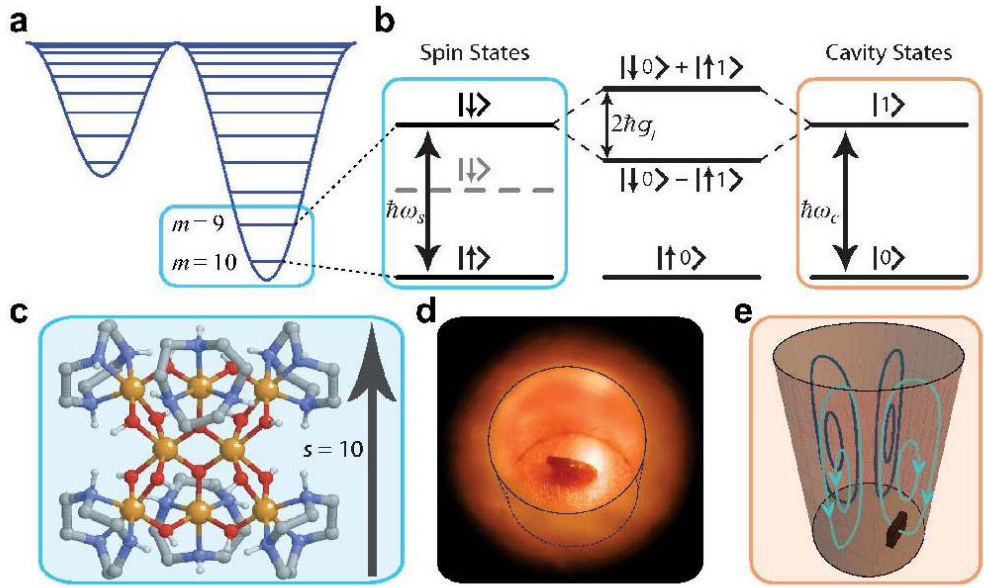

Fig. 1 a) Double-well potential for a single-molecule magnet. The levels correspond to different spin-orientation states. A magnetic field tilts the potential and increases the energy spacing between the levels in the lower well. b) Schematic of spin-photon interaction. One pair of SMM levels (the lowest two for the case shown) are labeled |↑> and |↓>, as shown. Photon-number states are labeled |0> and |1>. The photon energy $\hbar\omega_C$ for a given cavity mode is fixed. As the magnetic field is increased the energy between the two spin states $\hbar\omega_S$ increases, causing the excited state |↓> to shift upwards (from the grey, dashed level). When the energy of the |↓> state is near the |1> state, the two states hybridize, as shown in the middle of the panel. The splitting $2\hbar g_1$ (the smallest energy difference between the two levels) is determined by the strength of the interaction between the spin and photon systems. c) Structure of the $Fe_8$ SMM studied. It behaves as an anisotropic spin-10 system with a double-well potential similar to that in (a). d) Photograph of an $Fe_8$ single crystal mounted in a cylindrical resonant cavity. Some lines have been added to guide the eye. e) Structure of the $TE_{011}$ resonant mode excited in the cavity. Magnetic field lines associated with radiation in this mode are shown. The parallelepiped at the bottom of the cylinder is at the approximate position of the sample in the cavity (as shown in (d)).

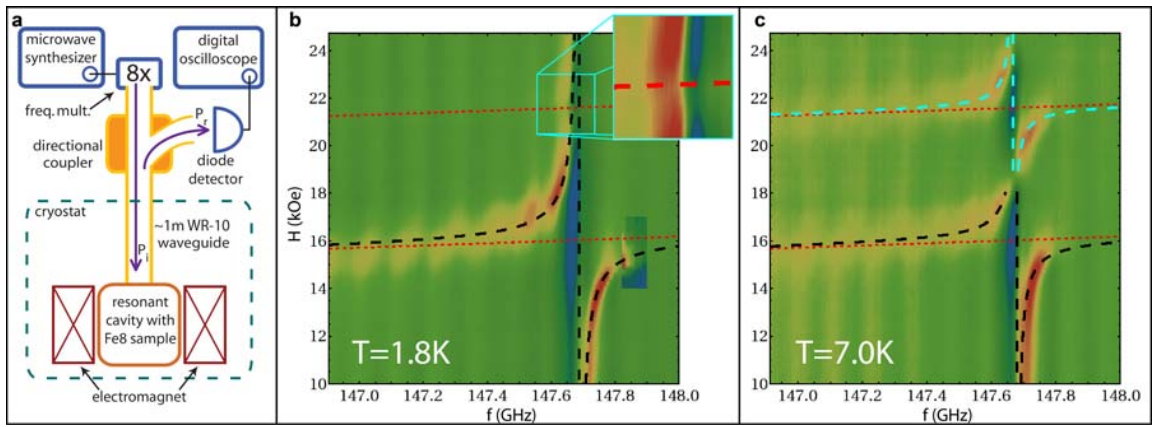

Fig. 2 a) Schematic diagram of experimental apparatus. b) and c) Absorbed power as a function of magnetic field and radiation frequency at 1.8 K and 7.0 K, respectively. Yellow and red indicate regions of significant power absorption by the sample-cavity system. The dark blue regions are largely artificial, produced by our background-subtraction procedure – see Supplementary Information. The lower (upper) red dashed line is the Zeeman energy separation for the $m$ = 10-to-9 (9-to-8) transition. The dashed black curve is a fit of the data for the lower-field data to Eq. 4. The cyan curve in **c** is a fit to Eq. 4 for the higher-field data, using a slightly smaller bare cavity frequency to account for remnant effects from the lower-field transition. The outset of (b) shows a zoomed-in view of the boxed region using a slightly different coloring scheme to enhance the feature associated with the 9-to-8 transition. Similarly, the data within the lower boxed region in (b) uses a different coloring scheme to emphasize the weaker features within the region. In both (b) and (c), an essentially field-independent background was subtracted to enhance visual presentation (see Supplementary Information). Remnants of this background appear as the modulation of signal with frequency. Analysis of the data treated the background using a more rigorous method (see Supplementary Information).

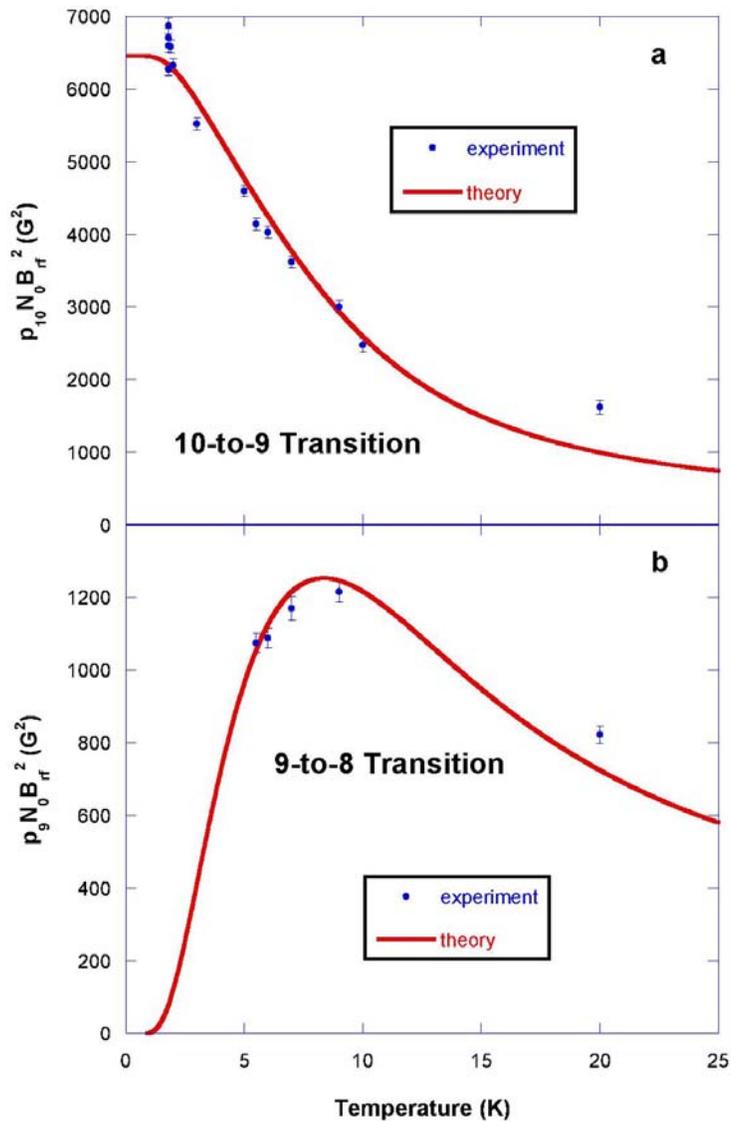

Fig. 3 Measured frequency splitting as a function of temperature for the (a) $m$ = 10-to-9 transition and (b) $m$ = 9-to-8 transition. The splitting has been recast in terms of relative level population $p$ of the lower level (see text) and compared with populations calculated using the known energy levels for the $Fe_8$ single-molecule magnet (solid curves). The vertical scaling factors needed to bring the calculated populations to coincide with the data yield values for the product $N_0 B_{rf}^2$.

## Supplementary Information

Apparatus and Data Acquisition

We used a high-frequency synthesizer followed by an 8x frequency multiplier chain to produce millimeter-wave radiation, which was transmitted down a ~ 1-meter waveguide to the cavity and sample. Radiation coupled to the cavity via a circular iris in the top plate of the cavity. The cavity was machined out of OFHC copper and had a radius of 1.28 mm and a depth of 3.96 mm. Measurements were performed in a Quantum Design PPMS. A directional coupler in conjunction with a diode detector enabled us to measure the reflected power as a function of the frequency of the incident power; dips in reflected power appeared at the resonant frequencies of the cavity-sample system. The reflected power as a function of frequency was recorded with a digital oscilloscope and the data was repeatedly uploaded to a computer as the magnetic field was slowly varied. The resulting data file for an experimental run contained data of reflected power as a function of radiation frequency and applied magnetic field. Experimental data was taken at several temperatures between 1.8 K and 20 K.

Data Analysis

The broadband detector employed in our apparatus contains an inverting diode, such that higher levels of power absorption (indicative of resonances of the system) appear as peaks in the data rather than as troughs. Thus, the data can be simply interpreted as representing absorbed power relative to an unimportant offset.

The upper panel in Fig. 1 shows raw data for absorbed power as a function of frequency and magnetic field at 7.0 K. The vertical bands in the figure are due to low-finesse resonances of our probe's waveguide, resulting in a roughly sinusoidal variation of power with frequency. While the signal from these background resonances varies significantly with frequency, it depends weakly on the applied magnetic field. It is straightforward to largely remove the background signal by taking the frequency dependence at low or high magnetic fields, i.e. far from where the sample interacts with the cavity, as reference data and subtracting it off from the data at all fields. Such an operation (followed by adjustments to color scale for visual clarity) performed on the 1.8 and 7.0 K data yields Figs. 2b and 2c (in the main text), respectively. The subtraction procedure results in the artificial dark blue regions in those figures because the procedure subtracts the bare cavity resonance from the data. We emphasize that these figures are produced for purposes of visual displaying the data. Actual data analysis was done on the raw data, as we explain presently.

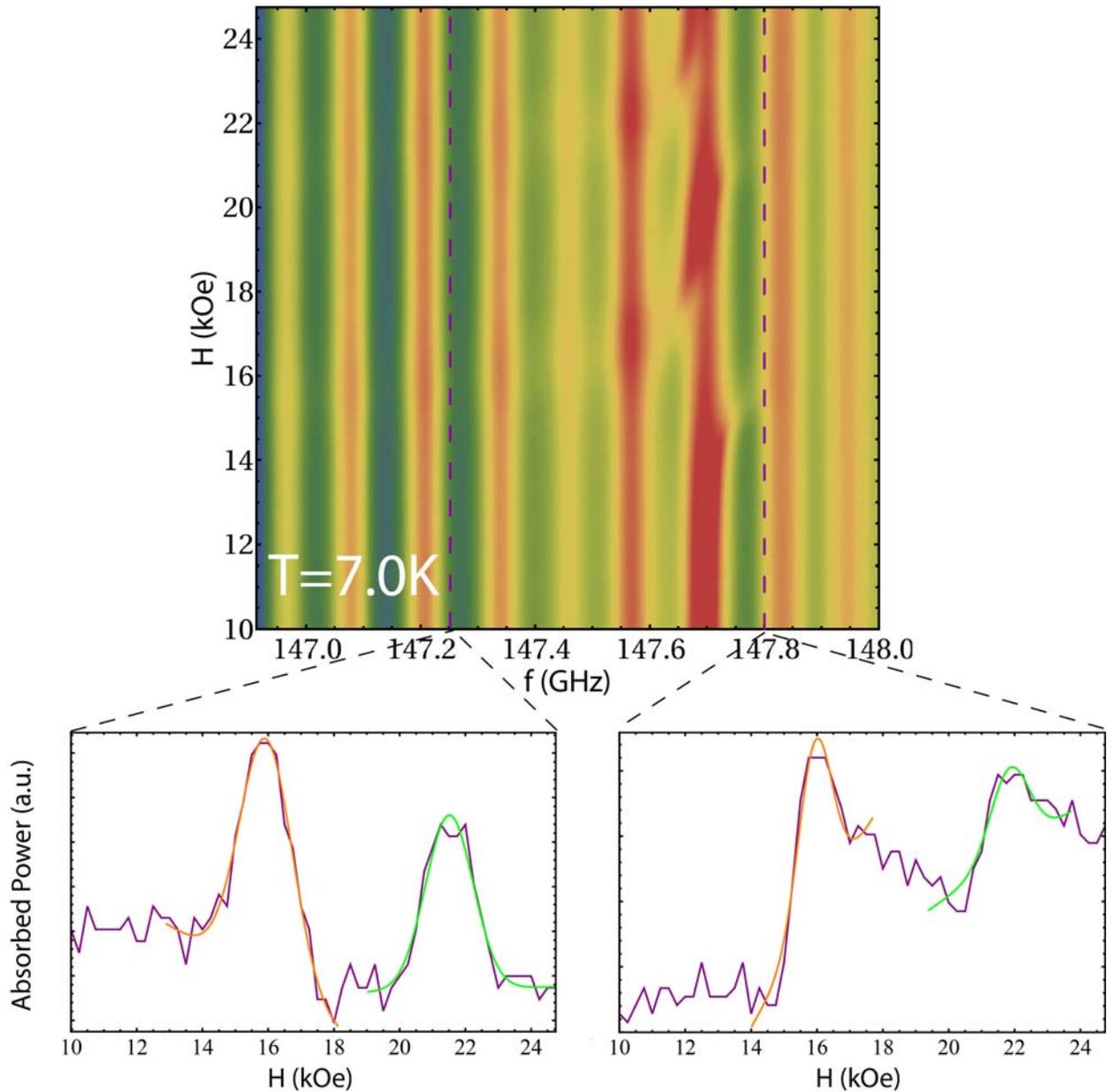

**Fig. 1.** Absorbed power as a function of frequency and magnetic field at T = 7.0 K. The lower panels show absorbed power as a function of magnetic field for a fixed frequency, i.e. data along the indicated vertical dashed line in the top panel. The orange and green curves are fits to a Gaussian plus a line, the latter to account for the behavior of the background. The vertical scales for the two lower panels are somewhat different.

Since the background is largely independent of magnetic field, we can isolate the spin resonance peaks from the background by analyzing constant-frequency subsets of the data. The lower panels of Fig. 1 show constant-frequency slices (along the thin vertical dashed lines in the upper panel) of absorbed power as a function of field. Two peaks are clearly visible, the lower-field one corresponding to the 10-to-9 transition and the higher-field one corresponding to the 9-to-8 transition. Some field dependence in the background is also observed. To find the position of each peak (magnetic field value for the resonance), we separately fit the data in the vicinity of each peak to a Gaussian plus a line, the latter to account for the variation of the background. Such a fit provides resonant-field values and uncertainties at each value of applied microwave frequency. The peak positions obtained in this way for one data set are shown by the orange and green points in the upper panel of Fig. 2. The fitting procedure returned a small number of clearly spurious data points, which were omitted from subsequent analysis. The oscillations in the peak positions as a function of frequency are remnants of background effects that were not fully accounted for by our fitting procedure. Because the magnitude of these background fluctuations is generally much larger than the uncertainties in the measured peak positions, we neglected these uncertainties in subsequent fitting.

We next fit the frequency dependence of the peak field positions. The expected dependence can be obtained from Eq. 4 (main text) and the very good approximation that the spin resonance frequency, $f_S = \omega_S/2\pi$, depends linearly on field: $f_S = f_0 + \beta H$. Making this substitution in Eq. 4 and solving for $H$ yields

$$H = \frac{1}{\beta}\left(\frac{(g_N/2\pi)^2}{f_C - f} + f - f_0\right), \tag{1}$$

where $f = \omega/2\pi$ is the resonance frequency of the coupled system and $f_C = \omega_C/2\pi$ is the bare-cavity resonance frequency. The constants $f_0$ and $\beta$ depend on the anisotropy parameters of the Hamiltonian and the angle $\theta$ between the easy axis and the magnetic field. $f_0$ and $\beta$ depend weakly on the azimuthal angle $\phi$, which determines the orientation of the field in relation to the intermediate (x) axis of the $Fe_8$ molecule. The dependence on $\phi$ is sufficiently weak that it is not a reliable fitting parameter and we simply fixed $\phi$ at the expected value of 108.7°, based on the crystal's orientation. We set the Hamiltonian parameters at the values given in the main text and let $\theta$ be a free parameter. $f_0(\theta)$ and $\beta(\theta)$ were then calculated by diagonalizing the Hamiltonian to obtain its eigenenergies, then determining the frequency $f_S$ for the relevant (e.g. 10-to-9) resonance as a function of $H$ and $\theta$, and fitting the $H$ dependence to a line in the experimentally relevant range of field values. Eq. 1 was simultaneously fit to the resonance peak positions corresponding to the relevant spin resonance with $\theta$, $f_C$ and $g_N$ as fitting parameters. $\theta$ was forced to be the same for each data set while $f_C$ and $g_N$ were allowed to vary from one data set to the next. Our fits yielded $\theta = 37.7°$ and a separate best-fit value of $g_N$ for each data set. The upper panel of Fig. 2 shows the peak-position data obtained from the data shown in Fig. 1 and the resulting fit to Eq. 1, following the procedures described above. The lower panel in Fig. 2 results from applying a similar procedure to data at 1.8 K. At that low temperature, the higher-field (9-to-8) resonance does not display an unambiguous splitting and only the signal from the lower-field (10-to-9) resonance was analyzed.

The values of $f_C$ obtained from our fits show a variation with temperature on the order of 10 MHz. $f_C$ values for the 9-to-8 resonance were generally somewhat smaller than those for the 10-to-9 resonance. This weak behavior is likely due to a combination of changes in mean dipolar fields with temperature and the remnant effect of one spin-cavity interaction on the other one. Neither of these effects is significant enough to substantially impact our main conclusions.

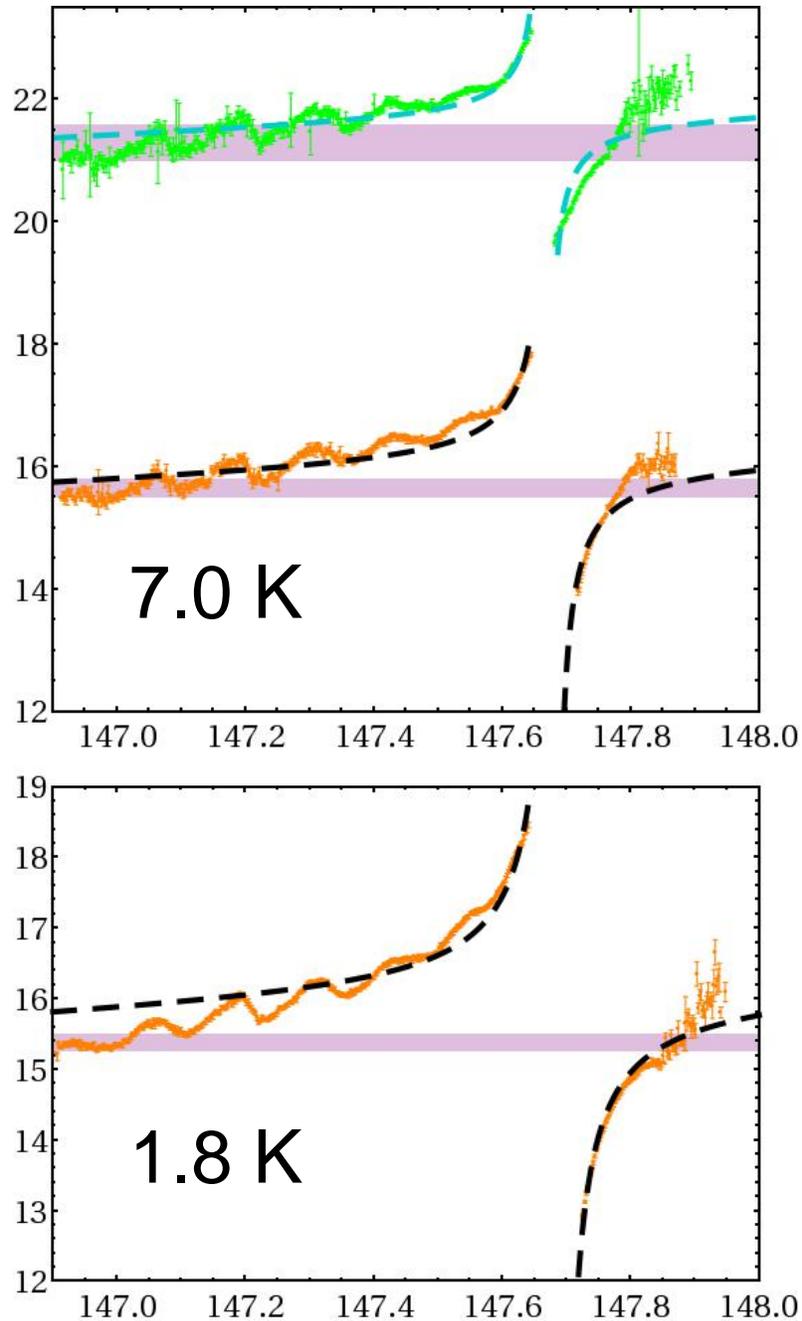

**Fig. 2. Resonance peak positions as a function of frequency (orange and green points) determined from fitting the data, as described in the text. Data for 7.0 K (i.e. obtained from an analysis of Fig. 1) and for 1.8 K are shown. The oscillations of the peak positions are due to remnant effects of the frequency-dependent background. Error bars represent standard errors. The data is fit (dashed lines) to Eq. 1. The shaded regions indicate the ranges of magnetic fields over which the system unambiguously exhibits two resonant frequencies.**

One important feature of the data is the fact that for some regions of field, the system has two distinct resonant frequencies, as illustrated in Fig. 2 by the shaded regions. This observation indicates that the spin-cavity system is in the so-called strong-coupling regime.

The values of $g_N$ obtained from our fitting procedure are plotted in Fig. 3 in the main text and fit to the energy-level populations with one adjustable parameter, as described in the main text. In our calculations, we also included levels in the known $s = 9$ spin manifold for $Fe_8$.[1]